\documentclass{amsart}

\usepackage{amsthm}
\usepackage{amssymb}
\usepackage{amsmath}

\newtheorem {Theorem}   {Theorem}

\numberwithin{Theorem}{section}

\newtheorem {Lemma}[Theorem]    {Lemma}

\theoremstyle{definition}
\newtheorem{Definition}[Theorem]{Definition}
\theoremstyle{remark}
\newtheorem{Remark}[Theorem]{Remark}

\newtheorem {Corollary}[Theorem]{Corollary}

\expandafter\chardef\csname pre amssym.def at\endcsname=\the\catcode`\@
\catcode`\@=11
\def\undefine#1{\let#1\undefined}
\def\newsymbol#1#2#3#4#5{\let\next@\relax
 \ifnum#2=\@ne\let\next@\msafam@\else
 \ifnum#2=\tw@\let\next@\msbfam@\fi\fi
 \mathchardef#1="#3\next@#4#5}   
\def\mathhexbox@#1#2#3{\relax
 \ifmmode\mathpalette{}{\m@th\mathchar"#1#2#3}%
 \else\leavevmode\hbox{$\m@th\mathchar"#1#2#3$}\fi}
\def\hexnumber@#1{\ifcase#1 0\or 1\or 2\or 3\or 4\or 5\or 6\or 7\or 8\or
 9\or A\or B\or C\or D\or E\or F\fi}

\font\teneufm=eufm10
\font\seveneufm=eufm7
\font\fiveeufm=eufm5
\newfam\eufmfam
\textfont\eufmfam=\teneufm
\scriptfont\eufmfam=\seveneufm
\scriptscriptfont\eufmfam=\fiveeufm

\catcode`\@=\csname pre amssym.def at\endcsname
 
\begin{document}






\title[The Motion of a Charged Particle]{The Motion of a Charged Particle
on a Riemannian Surface under a Non-Zero Magnetic Field}

\author[C\'esar Castilho]{C\'esar Castilho}
\address{Departamento de Matem\'atica, Universidade Federal de Pernambuco,
Recife, PE, CEP 50740-540, Brazil}

\email{castilho@dmat.ufpe.br}

\date{August, 1999}

\thanks{This work was supported by  CNPq-Brazil}

\bigskip

\begin{abstract}
In this paper we study the motion of a charged particle on a Riemmanian
surface under the influence of  a positive magnetic field $B$.
 Using Moser's Twist Theorem
and ideas from classical pertubation theory we find sufficient conditions
to perpetually trap the motion of a particle with a sufficient large
charge in
a neighborhood of a level set of the magnetic field. The conditions
on the level set of the magnetic field that guarantee the trapping are local
 and hold near all non-degenerate critical local minima or maxima of $B$ .
 Using sympletic reduction we apply the results of our work to certain
 $S^1$-invariant magnetic fields on ${\mathbb R}^3$.
\end{abstract}
\maketitle
\section{Introduction}
 
The motion
 of a charge $e $ on a Riemannian surface $M$  can be described as a
Hamiltonian dynamical system on $T^*M$ with 
metric Hamiltonian and  twisted symplectic 2-form $\omega_e = d\lambda
+ e \pi^* \beta$. Here $\lambda$ is the standard Liouville 1-form, $\pi : T^*M
\rightarrow M$ is the canonical projection, $e$ is the particle's charge and
$\beta$ is a  2-form on $M$ representing the magnetic field. A method
for finding periodic orbits of this motion
on a fixed energy level was introduced by V.I. Arnold (\cite{Ar2,Ar3}) and
developed by V.Ginzburg (\cite{Gi1,Gi2,Gi3}). The
method relies on symplectic topology techniques (Conley-Zehnder theorem 
\cite{CZ})or
symplectic capacities) to prove the existence of some minimum number of
periodic
orbits. In this work we will combine  ideas from classical
pertubation theory and appropriately chosen changes of varibles
(cf. \cite{Arn} and \cite{Gi2})
to study  motion when the charge is large. The approximation involved
with large charge is known as the guiding center approximation. The motion
of the particle is described as a fast rotation around a
point that drifts slowly along a level line of the magnetic field. The
point of rotation is known as the guiding center. Its behaviour is
important since one is usually interested in confining charged particles
to a bounded region of the phase space. We will find local
conditions on a level set of a magnetic field on a surface with general
Riemannian metric to perpetually trap the motion of the particle near this
level set.
 By considering the twisted symplectic
2-form as the main element of our treatment we will write  the system 
in a form suitable for applying Moser's Twist Map Theorem after
applying successive diffeomorphisms. The motion will be trapped between the
invariant circles predicted by the theorem.
The non-degeneracy condition of the theorem will be studied and neighborhoods
where the condition of the theorem holds will be characterized. Throughout the
work the
energy level will be fixed at $E=\frac{1}{2}$. We state our main results:
 
\begin{Theorem}
\label{ppric} Let $M$ be a Riemannian surface. Let $\Omega$ be the area form
on $M$. Let $ w = d\lambda + e\pi^*( B \Omega)$, where $\lambda$ is the
canonical 1-form on $T^*M$, $\pi^* : T^*M \rightarrow M$ is the natural
 projection, $B$ is a positive $C^2$ function on $M$  and $e > 0$ is a
parameter. Denote by $L_c=\left\{ B=c \right\}$ a non-critical level set of
 the function $B$ with the topology of an imbedded circle.
Set $\beta \equiv B\Omega = dA $. Let $F=\frac{ \nabla B}{ |\nabla B|^2}$
where $\nabla B$ denotes the gradient of $B$.
 If the level set $L_c$ is such that
\begin{equation}
\int_{L_c} i_F d \left\{ i_F dA + \frac{ 2}{ c} A \right\} \ne 0
\label{dedege}
\end{equation}
then for any neighborhood $N$ of $L_c$, there exists a number $ e_{*} > 0$
such  that for any $ e > e_{*} $  the trajectories of the magnetic field
problem with energy $E=\frac{1}{2}$ and charge $e$ will be contained on $N$
\underline{for all times}.
\end{Theorem}

\begin{Corollary} Let $p \in M$ be a non-degenerate maximum or minimum
point of
the function $B$. Then there exists arbitrarily small neighborhoods
$N$ of $p$, and $e_{*} > 0,$ such that
for all $e > e_{*}$ the integral curves of the magnetic problem with charge
$e$, energy $E=\frac{1}{2}$ and initial conditions in $N$ will remain in $N$
\underline{for all times}.
\label{nod1}
\end{Corollary}

\begin{Corollary} Let $s \in M$ be a  maximum or minimum submanifold of
the function $B$ with the topology of an imbedded circle. Then there
exists arbitrarily small neighborhoods $N$ of $s$, and $e_{*} > 0,$ such that
for all $e > e_{*}$ the integral curves of the magnetic problem with charge
$e$, energy $E=\frac{1}{2}$ and initial conditions in $N$ will remain in $N$
\underline{for all times}.
\label{nod2}
\end{Corollary}

  The proof of the theorem will be
based on a change of coordinates that will put the twisted symplectic
2-form $\epsilon d\lambda + B \Omega $ in the form $ \gamma -  \epsilon^2
dH_{\epsilon} \wedge d\theta $,
 where $ H_{\epsilon}$ is an analytic function depending on $\epsilon$
such that its first zero order term is given by $B^{-1}$ i.e,
we will construct a diffeomorphism $\Lambda$ such that $\Lambda^*(\epsilon
d\lambda +
 B \Omega)= \gamma -  \epsilon^2 dH_{\epsilon} \wedge d\theta $. \par
The characteristic
 line field of this form is spanned by $\epsilon^2 X_{H_{\epsilon}}
+ \frac{\partial}{\partial \theta}$ where $X_{H_{\epsilon}}$ is the
hamiltonian
vector field of $H_{\epsilon}$ with respect to the area form $\gamma$.
 Thinking of the variable $\theta$ as time 
 allows us to introduce action-angle
 variables $I$, $\psi$ on a neighborhood of the level set $L_c$ of $B$,
 and to reduce the system to a time $2\pi$ map
 $\theta \rightarrow \theta + 2\pi$ from this neighborhood
 to itself. The particular form of $H_{\epsilon}$ and $\gamma$ will allow
 us to invoke the twist theorem provided condition \ref{dedege} holds for
$B$.\par
 The paper is organized as follows. In section 2 we introduce the tools
we need and describe the limit of large charge as a pertubative
limit. In section 3 we construct the diffeomorphism $\Lambda$(theorem
\ref{pric2}).
In section 4 we apply Moser's twist theorem to our system and characterize
the non-degeneracy condition. In section 5 we prove theorem \ref{ppric} and
its
corollaries. In section 6 we apply our main theorem to a family of
symmetric
magnetic fields and finally in section 7 we prove some technical lemmas.

\subsection*{Acknowledgments.} I am grateful to Viktor Ginzburg
and Richard Montgomery for introducing me to this problem and for help
during the preparation of this work. It's also a
pleasure to thank Francesco Fass\'o and Ely Kerman for many discussions and
suggestions. I would like to thank CNPq-Brazil for financial
support. 

\section{The Magnetic Problem for Large Charge}
\label{secao2}
Let $M$ be a 2-dimensional oriented Riemannian manifold. Let
$\beta =B \Omega$ be the magnetic field 2-form, where $\Omega$ is the
area form on $M$ and $B$ is a non-vanishing function on $M$.
 The motion of a charge on the Riemannian surface $M$ can be identified
with a Hamiltonian flow (see \cite{Gi1}). Identifying the tangent and
cotangent
bundle by means of the Riemannian metric we denote by $H:TM \rightarrow
  R $ the energy function $H(X)=\frac{\parallel X \parallel^2}{2}$ where
$X \in TM$.
 Consider the Hamiltonian flow
of $H$ on $T^*M$ with respect to the twisted symplectic structure
$ w_e= d\lambda + e \pi^* \beta $. Here $\lambda=p_1 dq_1 + p_2 dq_2 $ is
the canonical 1-form
on $T^*M$, and thus $d\lambda$ is the standard symplectic 2-form,
$\pi^* :TM \simeq T^*M \rightarrow M$ is the natural projection and $e$ is the
charge of the particle. Henceforth $\beta$ and $ \pi^* \beta $ will
be identified.\par
 
 The twisted symplectic form $ w_e= d\lambda + e  \beta $ is not defined
 in the limit in which the charge is infinite. We will rescale the twisted
 2-form so as to have an analytic limit.  Let $\epsilon = \frac{1}{e}$.
Set
$$ w_{\epsilon} \equiv \frac{1}{e} w_e = \frac{1}{e} \left( d\lambda + e 
\beta \right)
= \epsilon d\lambda + \beta. $$
 Denoting by $X_H$ the hamiltonian vector field of the magnetic problem
we have
$$ dH = w_e(X_H,\cdot) = e w_{\epsilon}(X_H,\cdot)= w_{\epsilon}   
(eX_H,\cdot).$$
 It follows that under the scaling $X_H$ gets mapped to
$\frac{1}{\epsilon}X_H$.
This can be interpreted as a reparametrization of time. Thus
 if $q(t)$ is a solution for $X_H$, then $q(\frac{t}{\epsilon})$
is a solution  curve for $\frac{1}{\epsilon} X_H$. Therefore the scaling
changes
periodic orbits of period $T$ for $X_H$  to periodic orbits of period
 $\epsilon T$ for $\frac{1}{\epsilon} X_H $.
 We thus have two associated problems. Given $H$ we can consider the
 Hamiltonian problem given by the symplectic 2-form $w_e$ whose vector field
 we denoted by $X_H$ or we can consider the Hamiltonian problem given
 by the symplectic 2-form $w_{\epsilon}$ whose vector field we denote   
 by $X_{\epsilon}$.  From now on we will work
 only with $w_{\epsilon}$. Write $X_{\epsilon} =\frac{1}{\epsilon}X_H$.  
 For future reference we collect these facts on the
 table 1.1.
\vskip 0.5cm
\begin{table}[ht]
\centering
\begin{tabular} {|c|c|c|c|}
 \hline
$\epsilon =  1 / e $ & Expression & Vector Field & Time \\ \hline  
$w_e$ & $ d\lambda + e \beta $ & $X_H$ & T \\ \hline
$w_{\epsilon}$ &  $\epsilon d\lambda + \beta $ & $X_{\epsilon}= X_H /
\epsilon $
 &  $\epsilon T $ \\ \hline
\end{tabular}
\label{tata}
\end{table}
\centerline{TABLE 1.1 Relations between the scaled and non-scaled problem.}
\bigskip
\par
 The limit in which $e$ goes to infinity can be understood as a
 pertubative limit. A closer look at Hamilton's equations will clarify
this point. First we introduce some concepts necessary for our study.\par
Denote by $S^1M \subset TM$ the
circle bundle over $M$ with respect to the given Riemannian metric. Then
$S^1M=\left\{H= \frac{1}{2}\right\}$. Let $v$ be the geodesic vector field
of $M$ and
$v_{\perp}$ be the vector field on $S^1M$ perpendicular to $v$ (with respect
to the orientation of the manifold). Denote
by $\phi$ the unit vector field in the
direction of the fibers of the
bundle $\pi : S^1M \rightarrow M $ with direction defined by the orientation
of $M$. Thus, $v$, $v_{\perp}$ and
$\phi$ form a global moving frame for
$S^1M$. Denote by $v^*$, $v_{\perp}^*$ and $\phi^*$ the dual frame.
Since energy is preserved the
 magnetic Hamiltonian vector field $X_H$ at energy $E=\frac{1}{2}$ is a vector
field over
$S^1M$. The following lemma gives the structure equations for the circle
bundle.
\begin{Lemma}
The Lie-Brackets of $v$,$v_{\perp}$ and $\phi$ are given by
$$  [v,v_{\perp}]= K \phi ,\hskip 0.5cm  [\phi,v]=v_{\perp} ,\hskip 0.5cm 
[\phi, v_{\perp}]=-v ; $$
where $K$ is the Gaussian curvature of $M$.
\label{tyuy}
\end{Lemma}
\begin{proof} See \cite{KN}.\end{proof}
Dual to this lemma is
\begin{Lemma}
$$ d\phi^* = K v_{\perp}^* \wedge v^* ,\hskip 0.5cm dv_{\perp}^* = v^* \wedge
\phi^*,\hskip 0.5cm dv^* = \phi^* \wedge v_{\perp}^*.$$
\end{Lemma}

\noindent \underbar{\it Remark:} $\phi^*$ is the connection 1-form for
the Levi-Civita connection.

\begin{Lemma} Let $\lambda$ denote the canonical 1-form on $T^*M$.
Then $\lambda = v^*$. \end{Lemma}

\begin{proof} In Darboux coordinates 
$$ \lambda = \sum_i p_i dq^i = \sum_{ij} g_{ij} v^j dq^i = \left<v,\cdot \,
\right> = v^* \, .$$ \end{proof}

\begin{Corollary}
Let $\Omega = v^* \wedge  v_{\perp}^*$ be  the area form on $M$ pulled
back to $S^1M$,
$\lambda$ the canonical 1-form  and 
$v^*$, $v_{\perp}^*$ and $\phi^*$ as before. Then
$$ d \left( i_{v_{\perp}} \Omega \right) =    \phi^* \wedge  v_{\perp}^* =
 d \lambda . $$
\end{Corollary}

  Consider the Hamiltonian equation for $w_e$,
\begin{equation}
 w_e(X_H, \cdot \, ) = -dH.
\label{mmj}
\end{equation}
We write the Hamiltonian vector field as 
\begin{equation} X_H = v + e f  \phi,\label{ham} \end{equation}
i.e, as a geodesic component  plus a fiber 
component. The function $f$ is calculated by inserting this
decomposition in equation \ref{mmj},
$$ \left( d\lambda + e B \Omega \right) \left( v + e f \phi  \right) = dH. $$
\noindent Since $\Omega  \phi = 0 $ ,
$d\lambda (v) = dH = 0$ (on $S^1M$),
$\Omega (v) = v_{\perp}^*$ and $d\lambda (\phi) = v_{\perp}^* $.
We find that
$f=-B$. Thus $$ X_{H} = v - e B \phi. $$
in $S^1M$. Rescaling yields
\begin{equation}
 \frac{X_H}{e} = \epsilon v - B \phi.  
\label{arere}
\end{equation}
Set $\tilde X = X / eB$ and  $ w =\frac{v}{B}$. Dividing
 by $B$
  we have that
 $$ \tilde X  = \epsilon w - \phi.$$
 The integral curves of $X_{\epsilon}$ and $X$ are the same up to
reparametrization. Thus the limit
$0 < \epsilon << 1 $
can be understood as the limit where the vertical vector field $- \phi$
is perturbed by a `small' vector field $ \epsilon w $. \par
This interpretation allow us to apply standard pertubation theory
to eliminate, in the first approximation, the pertubation containing $w$
by means of an appropriate change of coordinates. We
indicate how to make this change of coordinates.
Let
$$ x \prime = X (x)= \epsilon w(x) - \phi(x),$$
where  $\prime$ denotes derivation with respect to the new time.
 The basic idea is to pull
back the vector field $x\prime$ by the $\epsilon $-time flow
$\Phi_{\epsilon}$ of some vector field $Y$.
For a small $\epsilon$ we have (formally);
  $$ \Phi_{\epsilon}^* \left( x\prime \right) = \sum_{s=0}^{\infty}
\frac{\epsilon^s}{s!}\left.\frac{d^s}{d \tau^s}\right|_{\tau=0}\Phi_{\epsilon} 
 \left( x\prime \right),  $$
 where
$\tau= \frac{t}{\epsilon}$. \noindent But 
$$\left. \frac{ d}{d \tau}\right|_{\tau =0} \Phi_{\epsilon}^* x\prime = 
[ Y , x \prime ]  ;$$

$$\left. \frac{ d^2}{d \tau^2 }\right|_{\tau =0}\Phi_{\epsilon}^* x\prime =
   [ Y, [ Y , x\prime ] ];$$
$$\left.\frac{ d^3}{d \tau^3 }\right|_{\tau =0}\Phi_{\epsilon}^* x\prime =
 [Y , [Y, [ Y , x\prime ] ]] ,$$
and so on. Since $ x \prime = \epsilon w - \phi $ the first order term of
 $ \Phi_{\epsilon}^* \left( x\prime \right)$  is given by (recall that  
$ w = \frac{v}{B}$):
$$  \frac{ v}{B} -  [ Y , \phi].$$
\noindent The goal is to choose $Y$ such that the first order term of
 the expansion is zero, i.e.
\begin{equation}  \frac{ v}{ B} -  [ Y ,\phi] = 0. \label{eik}  \end{equation}
\noindent In classical pertubation theory, equation \ref{eik} is called
the eikonal equation (see \cite{Arn}).
\begin{Lemma} The vector field $v_{\perp} / B$ solves the eikonal
equation \ref{eik}. \end{Lemma}
\begin{proof} This follows directly from lemma \ref{tyuy}:
$$ \frac{v}{B} - [ \frac{v_{\perp}}{B }, \phi] = \frac{v}{B} + \frac{ 1}{B} 
 [\phi , v_{\perp}]=\frac{v}{ B} - \frac{ v}{ B} = 0. $$ \end{proof}

\section{The Pertubative Approach}
\label{secao3}
 The flow of the vector field ${v_{\perp}}/{B }$ gives the
diffeomorphism that eliminates the first order term in the $\epsilon$
expansion. In this section we prove the following theorem:
\begin{Theorem} 
\label{pric2}
Assume $L_c= \left\{ B=c \right\}$ is a simple closed curve.
 For $ \epsilon $ sufficiently small ($\epsilon > 0$), there exists a
neighborhood $N_c$ of $\pi^{-1} \left(L_c \right) \subset S^1M$
(depending on $\epsilon$) and  a diffeomorphism
$\Lambda :N_c \rightarrow N_c$  such that $$ \Lambda^* {w_{\epsilon}} =
\gamma - (\epsilon^2 dB^{-1} + \epsilon^3 dF) \wedge d\theta ,$$
where $\gamma$ is the area form on $M$, $F$ is a function on $S^1M$
and $\theta$ is a circular fiber coordinate on $S^1M$.
\label{pronn}
\end{Theorem}
 The change of coordinates $\Lambda$ is involved. The basic idea  is to
construct
 it first on a section
 $\theta = constant $ of the circle bundle and then extend it to the full
 circle bundle. The diffeomorphism on this section  is built by pulling
 back the twisted 2-form by the flow of
 the vector field ${ v_{\perp} / B}$ introduced in section \ref{secao2}.
In the flat case the triviality of the bundle implies
that this is
 already sufficient to allow the application
 of the twist theorem. In the non-flat case we will need one diffeomorphism
 for each such section of the principal bundle. This apparent
difficulty will be overcome by  `gluing' together all these diffeomorphims.
The main
 tool used to make this gluing
 will be a version of Moser's homotopy argument which says that two
symplectic
 2-forms that agree on a compact submanifold must agree on a
neighborhood of the submanifold. The resulting diffeomorphism, now
 defined on the full bundle, will put the twisted symplectic 2-form in the
final desired form.\par
 
\begin{proof}
   We start to build the diffeomorphism. We compute the pull-back of the
 twisted symplectic 2-form by the flow of  ${ v_{\perp} / B}$. The choice
of this vector field is dictated by its role  in the pertubation problem
 as shown on the previous section.

\begin{Lemma} The pull-back of $ w_\epsilon $ by ${\varphi_\epsilon}$, the
$\epsilon$-time flow of $v_{\perp} / B $
is given by
\begin{equation}
{\varphi_\epsilon}^* w_\epsilon= B\Omega + \frac{\epsilon^2}{2} \frac{K}{B}
\Omega - \frac{\epsilon^2}{2} dB^{-1}\wedge \left( d\theta + f \right) +
 O(\epsilon^3),
\label{expre}
\end{equation}
where $\theta $ is the fiber coordinate function and $f$ is a 
fiber-independent
 1-form.
\end{Lemma}
\begin{proof}
This is a straightforward computation using Cartan's Formula and the
structure equations. The identity $\phi^*=d\theta + f$ follows
from the fact that $\phi^*$ is the Levi-Civita connection
restricted to $S^1M$. \end{proof}
\begin{Corollary} {\bf The Flat Case :}  Suppose $M$ is flat. Then  $ K = 0
$, the bundle is trivial
and we can take $\phi^*=d\theta$ i.e. $f=0$. The pull-back
computations reduce to
$$ L_{v^\perp / B} w_{\epsilon} = -d\lambda - \epsilon
dB^{-1} \wedge
d\theta .$$
$$ L_{v^\perp / B}^2 w_{\epsilon} = \left( dB^{-1} - \epsilon
 L_{v^\perp / B} dB^{-1}
\right) \wedge d\theta . $$
$$L_{v^\perp / B}^3 w_{\epsilon} = \left( L_{v^\perp / B}
dB^{-1}-\epsilon
L_{v^\perp / B}^2 dB^{-1} \right) \wedge d\theta. $$
We obtain 
\begin{eqnarray} {\varphi_\epsilon}^* w_\epsilon & = &
\epsilon d\lambda + B\Omega + \epsilon
 \left(-d\lambda - \epsilon  dB^{-1} \wedge d\theta \right) +
\frac{\epsilon^2 }{2} \left( dB^{-1} - \epsilon
 L_{v^\perp / B} dB^{-1}\right) \wedge d\theta + \nonumber \\
 & & \frac{\epsilon^3 }{3} \left( L_{v^\perp / B} dB^{-1}
-\epsilon
L_{v^\perp / B}^2 dB^{-1} \right) \wedge d\theta + \ldots
\end{eqnarray} 
which simplifies to
$$ {\varphi_\epsilon}^* w_\epsilon= B\Omega - \frac{\epsilon^2}{2}
 dB^{-1}\wedge d\theta - \frac{\epsilon^3}{6}L_{v^\perp / B}dB^{-1} \wedge
d\theta
- \frac{\epsilon^4}{12}L_{v^\perp / B}^2 dB^{-1} \wedge
d\theta  + \ldots $$
giving 

$$ {\varphi_\epsilon}^* w_\epsilon= B\Omega - \epsilon^2 dH_{\epsilon} \wedge
 d\theta ,$$
where
$$ H_{\epsilon} = \sum_{i=1}^{\infty} \frac{ \epsilon^{i-1}}{i(i + 1)}
L_{v^\perp / B}^{(i-1)} B^{-1}.
$$
 So, for the planar case, the first diffeomorphism is already sufficient
to bring the twisted symplectic 2-form to the desired form. One
should compared this expression with the one found by Littlejohn
\cite{Little}.

 \end{Corollary}
   A second diffeomorphism will be necessary if $K \ne 0$. 
  Denote $\bar w_{\epsilon } ={\varphi_\epsilon }^* w_{\epsilon }$. First
 we will restrict $\bar w_{\epsilon }$ to the level set $\theta = constant$.
 In this case we have that $d\theta = 0 $ and (\ref{expre}) becomes 
\begin{equation}
\bar w_{\epsilon } =  B\Omega + \frac{\epsilon^2}{2} \frac{K}{B}
\Omega - \frac{\epsilon^2}{2} dB^{-1}\wedge  f  + \dots
\label{conn}
\end{equation}

\begin{Lemma}
 For a constant $\theta$ and a sufficiently small $\epsilon$ ($\epsilon > 0$)
 there exist neighborhoods $N_o$ and $N_1$ (depending on $\epsilon$) of
$\pi^{-1}\! \left(L_c\right) \subset S^1M$, and a diffeomorphism
$\Phi_{\epsilon , \theta} : N_0 \rightarrow N_1$
such that $\Phi_{\epsilon , \theta}^* \bar{w_{\epsilon}} = B\Omega $. Moreover
$$ \Phi_{\epsilon , \theta}= Id + \epsilon^2 C(\epsilon)\label{lema1} $$  
with $C(\epsilon)$ uniformly bounded in the $C^0$ norm.
\label{limee}
\end{Lemma}
\begin{proof}
 $\pi^{-1}(L_c) \cap \left\{ \theta =constant \right\}$ has dimension 1,
 consequently the forms $\bar w_{\epsilon}$ 
and $B\Omega$
degenerate. Since $L_c$ is compact it follows from
theorem of Weinstein (a version of Moser's homotopy argument)
that there exists a diffeomorphism $\Phi_{\epsilon,\theta}$ such that
$\Phi_{\epsilon,\theta}^* \bar w_{\epsilon} = B\Omega$. The order of 
the diffeomorphism is given by tracing back Moser's homotopy argument. This is
done in section \ref{append} \end{proof}
 Composing the two
 diffeomorphims $\Phi_{\epsilon,\theta}$ and $\varphi_{\epsilon}$ we obtain
 on the section  $\theta = constant $
 of the principal bundle that
$$\left( \varphi_{\epsilon} \circ \Phi_{\epsilon ,\theta}\right)^*
w_{\epsilon} =  \Phi_{\epsilon ,\theta}^* \circ  \varphi_{\epsilon}^* (w_
{\epsilon}) =  \Phi_{\epsilon ,\theta}^* \bar w_{\epsilon} = B \Omega . $$
 For a  fixed $\epsilon$, $\Phi_{\epsilon ,\theta} $ is a
 family of diffeomorphisms parametrized by $\theta $, each defined only on
 its own constant section of the principal bundle. We will paste all those
diffeomorphisms together. Let $p \in S^1M$. With respect to our local
trivialization we write $p=(m,\theta)$ for an arbitrary point of
$S^1M$ in the given neighborhood.
Define $\Xi_{\epsilon} : S^1M \rightarrow S^1M$ by
$$ \Xi_{\epsilon}(m,\theta) = \left( \Phi_{\epsilon ,\theta}(m), \theta
\right).$$

\begin{Lemma} Assume $L_c= \left\{ B=c \right\}$ is a simple closed curve.
 For $ \epsilon $ sufficiently small ($\epsilon > 0$), there exists a
neighborhood $N_c$ of $L_c$ 
 (depending on $\epsilon$) and  a diffeomorphism
$\Xi :N_c \rightarrow N_c$  such that $$ \Xi^* {w_{\epsilon}} =
B\Omega - (\epsilon^2 dB^{-1} + \epsilon^3 R) \wedge d\theta ,$$
for some 1-form R.
 
\label{prinn}
\end{Lemma}
\begin{proof} Since $\Xi_{\epsilon}$ restricted to $\theta=constant$ is
equal to
$\Phi_{\epsilon , \theta}$ if follows that $\Xi_{\epsilon}^*\bar w_{\epsilon}
$ differs
from $\Phi_{\epsilon , \theta}^* \bar w_{\epsilon}$ by a factor of the
form $\alpha \wedge d\theta $, for some 1-form $\alpha$, i.e., we can
write 
\begin{equation}
 \Xi_{\epsilon}^*\bar w_{\epsilon} = B\Omega + \alpha \wedge
d\theta .
 \label{iniii} \end{equation}
The one-form $\alpha$ is defined mod $d\theta$ and can be taken
to be
$$\alpha =- i_\frac{ \partial}{\partial \theta}  \Xi_{\epsilon}^*
 w_{\epsilon}. $$
For any diffeomorphism $\delta$ we have that $\delta_* \left( i_X \alpha
\right)= i_{\delta_* X} \left(\delta_* \alpha\right) $ (see e.g.
\cite{Foundations}). Therefore
\begin{equation}  i_\frac{ \partial}{\partial \theta} \Xi_{\epsilon}^*
\bar w_{\epsilon}  = \Xi_{\epsilon}^* (i_{\Xi_{\epsilon}*
 \frac{ \partial}{\partial \theta}}\bar w_{\epsilon}) .
\label{epp}
\end{equation}
So we must calculate ${\Xi_{\epsilon}}_*  \frac{ \partial}{\partial \theta}$.
 We do this on the following lemma.

\begin{Lemma} 
$$  \Xi_{\epsilon *} \frac{\partial}{\partial \theta} =
\frac{\partial}{\partial \theta}
 + \epsilon^3 z $$
where $z$ is a horizontal vector field relative to our trivialization.
\label{scct}  
\end{Lemma}
\begin{proof} The proof of this lemma is given in section \ref{append}
\end{proof}
 Using this lemma we compute (\ref{epp}) ,
\begin{equation}
i_\frac{\partial}{\partial \theta} \Xi_{\epsilon}^* \bar w_{\epsilon}=
\Xi_{\epsilon}^*(i_{\Xi_{\epsilon *} \frac{\partial}{\partial \theta}} \bar
w_{\epsilon})
=  \Xi_{\epsilon}^*(i_\frac{\partial}{\partial \theta}\bar w_{\epsilon}) +
\epsilon^3  \Xi-{\epsilon}^*(i_z \bar w_{\epsilon}).   
\label{zza}
\end{equation}
We recall the expression for $\bar w_{\epsilon}$ (formula \ref{expre})
$$\bar w_{\epsilon} \equiv {\varphi_\epsilon}^* w_\epsilon= B\Omega +
 \frac{\epsilon^2}{2} \frac{K}{B}   
\Omega - \frac{\epsilon^2}{2} dB^{-1}\wedge \left( d\theta + f \right) + 
O(\epsilon^3).$$
It follows that
$$i_\frac{\partial}{\partial \theta} \bar w_{\epsilon} = -\frac{
\epsilon^2}{2}
 dB^{-1} - \epsilon^3 G ,$$
 for some 1-form $G$. We have that \ref{zza} can be writen as
$$i_\frac{\partial}{\partial \theta} \Xi_{\epsilon}^* \bar
w_{\epsilon}=\Xi_{\epsilon}^*\left(
-\frac{ \epsilon^2}{2} dB^{-1} - \epsilon^3 G \right) +\epsilon^3
\Xi_{\epsilon}^*(i_z \bar w_{\epsilon}).$$
Using lemma \ref{limee} we have 
$$i_\frac{\partial}{\partial \theta} \bar w_{\epsilon} =(1_d + \epsilon^2
C^*(\epsilon,\theta))(-\frac{ \epsilon^2}{2} dB^{-1} + \epsilon^3 G)
+\epsilon^3 \Xi^*(i_z \bar w_{\epsilon}),$$
giving  
$$i_\frac{\partial}{\partial \theta} \bar w_{\epsilon} =-\frac{\epsilon^2}{2}
dB^{-1} - \epsilon^3 R,$$
for some 1-form $R$.
Finally this implies 
\begin{equation}
 \Xi_{\epsilon}^* \bar{w_{\epsilon}} = B\Omega + \alpha \wedge d\theta =
 B\Omega - (\epsilon^2 dB^{-1} + \epsilon^3 R) \wedge d\theta
\label{tetw}
\end{equation}
which proves lemma \ref{prinn}. \end{proof}

The 2-form (\ref{tetw}) is almost in the final form. The 1-form $R$
may not be necessarily exact. A final change of coordinates is needed.
So far we have been working on a neighborhood
of a constant level curve $\left\{ B=c \right\}$ of the magnetic field. 
By hypothesis this has the topology of an annulus. We introduce Fermi
coordinates $(x,y)$ on this annulus (see e.g. \cite{Mont}). The Fermi
coordinates parametrize the $L_c$ neighborhood $N_0$ on the
following way: Given a point $m \in N_0$, consider the segment of geodesic
$S$
that connectes $m$ to $L_c$ and has minimal length. Let  $y$ denote its
oriented length. Denote by $x$ the oriented length from
some arbitraty point $x_0$ in $L_c$ to the intersection of $S$ with $L_c$.
 Using these coordinates we define
 $$
a(\theta)= \frac{ 1}{ l(c)}\int_{B=c}   
R dx ,$$ where $l(c)$ denotes the length of the level set $L_c$.
\begin{Lemma} With $a(\theta)$ as above there exist functions $F$ and $g$
such that
$$ R = dF + a(\theta) dx + g d\theta .$$
\label{acco}
\end{Lemma} 
\begin{proof} Since $B$ is $\theta$ independent and
$ \Xi_{\epsilon}^* \bar{w_{\epsilon}}$ is closed (\ref{tetw}) implies that
\begin{equation}
dR \wedge d\theta = 0.   
\label{trey}
\end{equation} Let
$$R = R_x dx + R_y dy + R_{\theta} d\theta,  $$ and denote by
$R^*$
the restriction of $R$ to a constant section of $S^1M$, i.e.
$$ R^* = R |_{\theta = c}= R_x(x,y,c) dx + R_y(x,y,c) dy. $$
Thus (\ref{trey}) implies that
$$dR \wedge d\theta =\left( \frac{\partial R_y}{\partial x} -
\frac{\partial R_x}{\partial x} \right)dx \wedge dy \wedge d\theta = 0,$$
from which we see that $R^*$ is closed. Now define a
1-form $\xi$ on the section $\theta = c$ by
\begin{equation}
 \xi = R^*- a(c) dx.
\label{cos}
\end{equation}
Thus
\begin{equation}
 d \xi = 0 \qquad and \qquad  \int_{L_c} \xi  = 0,
\label{tur} 
\end{equation}
giving that
$$ a(c)=\int_{L^{\prime}} R^*$$
for any $L^{\prime}$ homotopic to $L_c$ (since $R^*$ is closed).
Since the section $\left\{ \theta = constant \right\}$ has the homotopy
type
of an annulus we have that
 equations \ref{tur} together imply that
$$ \xi = dF $$
for some function F.
Thus we can write \ref{cos} as
$$R^* = dF + a(c) dx. $$
But $R^*$ difers from $R$ by a factor of the form $g d\theta$ for
some function $g$ and we can write that
$$ R = dF + a(\theta) dx + g d\theta,$$
proving the lemma. \end{proof} 
Using lemma \ref{acco} we can rewrite \ref{tetw} obtaining that
\begin{equation} \Xi_{\epsilon}^* \bar w_{\epsilon} = B\Omega - (\epsilon^2
dB^{-1} +
\epsilon^3 dF +\epsilon^3 a(\theta)dx) \wedge d\theta . \label{cmes}
\end{equation}
Define a new variable 
$$ \bar y =  \int_0^y B dy.  $$
and the diffeomorphism
\begin{equation} \left\{ \begin{array}{l}
           X = x,\\ Y = \bar y - \epsilon^3 \int_0^{\theta} a(\theta)
d\theta , \\ \Theta = \theta. 
          \end{array} \right. 
\label{nnn1}
\end{equation}
Substituting back in \ref{cmes}, and denoting by ${\bf  w_{\epsilon}}$ the
pull-back of $\bar{w_{\epsilon}} $ by  $\Xi_{\epsilon}^* $ we obtain that
$$ {\bf w}_{\epsilon}   =   dX \wedge d Y +\epsilon^3 a(\Theta)dX
\wedge  d\Theta -(\epsilon^2 dB^{-1} +  \epsilon^3 dF+\epsilon^3 a(\Theta)dX)
\wedge d\Theta ,$$
i.e, 
$$ {\bf w}_{\epsilon} =  dX \wedge dY -(\epsilon^2 dB^{-1} +
 \epsilon^3 dF ) \wedge d\Theta , $$
that we write as
\begin{equation}  {\bf w}_{\epsilon} =  dX \wedge dY +\epsilon^2 dH_{\epsilon}
\wedge d\Theta , \label{yyy} \end{equation}
where $ H_{\epsilon} = B^{-1}+\epsilon F$
which proves theorem \ref{pronn} \end{proof}

\section{Moser's Twist Theorem}
          
\label{actanle}
 The characteristic 
 line field of (\ref{yyy}) can then be thought
as being generated by the Hamiltonian system given by the Hamiltonian
$H_{\epsilon}$ and symplectic 2-form $dX \wedge dY$. We introduce
action angle variables  $I$, $\psi$ on a neighborhood of the
level set $L_c$ of $B$. Using the Fermi coordinates 
(see section\ref{secao3})  we have that the level set is given by 
$L_{c}=\left\{  y=0,\theta
=0 \right\}$ and is contained in an open set
$ N=\left\{ -\delta \le y \le \delta , \theta = 0 \right\}$
for some $\delta \ge 0$ and $N \subset N_0$. Denote by ${\bf N}$ the image of
$N$ under
the diffeomorphism. Thus we have that $N$ becomes ${\bf N}=\left\{
-\frac{\delta}{B} \le Y \le \frac{\delta}{B} , \Theta = 0 \right\}.$
 Define
\begin{equation} I(X,Y)= \int_{A(X,Y)} dX\wedge d\bar Y. \label{acao}
\end{equation}
where $A(X,Y)$ is the annulus $ Y \le \bar Y \le 0 $ or $ 0 \le \bar Y \le
Y $.
Calling $\psi$ the variable canonicaly conjugate to $I$ we have that
$$ H_{\epsilon}= \frac{\epsilon^2}{ 2} B^{-1}(I) + O(\epsilon^3), $$
$$ dX\wedge dY = dI \wedge d\psi . $$
We consider the Hamiltonian equations with $ H_{\epsilon}$ on $A(X,Y)$. By
integrating and taking the time $\left( \theta = 2 \pi \right)$ flow we get
\begin{equation}
 \left\{ \begin{array}{l} I_1 = I + O(\epsilon^3),  \\
         \psi_1= \psi + \pi \epsilon^2 \frac{\partial}{\partial I} B^{-1}(I)
 + O(\epsilon^3). \end{array} \right. \label{action} \end{equation}
At this point we recall the statement of Moser's Twist map theorem:
\begin{Theorem} {\it (Moser, 1962 \cite{Mos})} Let $\Phi$ be a measure
preserving map
$\Phi(R,\theta)=(R_1,\theta_1)$ given by

\begin{equation} \left\{ \begin{array}{l} 
           R_1 = R + \epsilon^{\alpha}f(R,\theta,\epsilon), \\
         \theta_1 = \theta + \alpha + \epsilon^k \gamma(R) +\epsilon^l
g(R,\theta,\epsilon), \end{array} \right. \end{equation}
where $ \gamma\prime (R) \ne 0$, $k \le l$, f and g bounded. Then for
$\epsilon$ sufficiently small there exists an invariant curve $\Gamma$
surrounding $R=1$. More precisely there exists a differentiable closed
curve   
\begin{equation}
\left\{ \begin{array}{l}  R = F(\phi,\epsilon), \\
         \theta=\phi + G(\phi,\epsilon) \end{array} \right. \end{equation}
with $F,G$ of period $2 \pi$ in $\phi$ which is invariant under the map
$\Phi$.
\label{twist}
\end{Theorem}

To apply the twist map theorem \ref{twist} to \ref{action} it is sufficient
that $\frac{\partial^2 }{\partial I^2 } B^{-1}(I)  \ne 0 $ at $L_c$.
This
is the so called non-degeneracy condition and plays an essential role in
our problem. In what follows we will characterize the non-degeneracy
condition.
 
\subsection{The non-degeneracy condition}
\begin{Definition} A level set $L_c=\left\{ B=c \right\}$ is called 
non-degenerate if 
\begin{equation}
\frac{\partial^2 B^{-1}(I)}{\partial I^2} \ne 0 ,
\label{dege}
\end{equation}
along $L_c$.
\end{Definition}
Using the chain rule and assuming $L_c$ is a non-critical level
set of $B$ we can write that (\ref{dege}) is equivalent to
\begin{equation}
\frac{d^2 I }{dB^2} = -\frac{ 2}{ B} \frac{ d I}{ dB}.
\label{imag}
\end{equation}
Form (\ref{imag}) is particularly useful since one can explicitly
calculate
the derivative of $I$ with respect to $B$. To do so
we first note that $I$, the action variable, can be easily
expressed in terms of the Fermi coordinates $x$ and $y$. Since
$$ I(X,Y)=\int_{A(X,Y)} dX\wedge d\bar Y, $$
we have that
$$ I(x,y)=\int_{\bar A(x,y)}B dx \wedge dy.$$
where $\bar A(x,y)$ is the preimage of $A(X,Y)$ under the diffeomorphism
(\ref{nnn1}).
\begin{Lemma}
\begin{equation}
\frac{\partial I}{\partial B}=\int_{L_c} i_F dA  =\int_{L_c} \left( \frac{ B}{
|\nabla B| }
\right) ds,
\end{equation}
where $ds$ is the arc length of $L_c$.
\label{primeira}
\end{Lemma}
\begin{proof} Choose a vector potential $A = A_x dx + A_y dy$ for the
2-form $B dx \wedge dy$. Thus we have by Stokes theorem that
$$ I(B)= \int_{L_c} A. $$
Now consider the vector field $F = \frac{\nabla B }{ | \nabla B |^2 }$. Let
 $\phi_t$ be its time $t$ flow. Thus
$$\frac{d}{dt}(B(\phi_t(x,y))=<\nabla B, F> = 1, $$
implying that $\phi_{\epsilon}^*B = B + \epsilon.$ We can write that
$$ I(B + \epsilon ) = \int_{\phi_{\epsilon} (L_c)} A = \int_{L_c} \phi^*_{\epsilon}A.$$
Differentiating with respect to $\epsilon$ and using Cartan's formula we
obtain that
$$ \frac{ d}{dI} I(B+\epsilon)= \int_{L_c}(di_FA + i_FdA).$$
The first term on the integration vanishes since $di_FA$ integrates to zero
 along the closed curve $L_c$. To get the other term, observe that for any
smooth simple closed curve with smooth normal $n$ we have
$i_n dx \wedge dy = ds$ the arc length.
Since $F= \frac{ 1 }{| \nabla B | }n $ and $dA = B dx \wedge dy$ we find that
$i_FdA = \frac{ B }{ | \nabla B | } ds$ as desired. \end{proof} \par
\begin{Corollary}
\begin{equation}
\frac{\partial^2 I}{\partial B^2} = \int_{L_c} i_F d( i_F dA) .
\end{equation}
\label{segunda}
\end{Corollary}
\begin{proof}
Since the second derivative is expressed as
the integral of a 1-form, the same reasoning applied to compute the first
derivative can be applied to compute the second.\end{proof} 
The non-degeneracy condition can be written in a more explicit way by noting
that $ds$ can be written as $f(x,y)dx$ for some function $f > 0$. A long
and straightforward computation give us that the level $L_c$ is nondegenerate
if 
\begin{equation} \int_{L_c} \left\{  f \left(
\frac{3 }{|\nabla B|}
  - \frac{ B}{|\nabla B|^4} < \nabla |\nabla B|, \nabla B > \right) 
 +  \frac{ B}{ |\nabla B|^3 } < \nabla f , \nabla B > \right\} dx \ne 0,
\label{deg}
\end{equation}
This expression for the non-degeneracy condition will be used to prove 
corollary \ref{nod1}.

\section{Proof of Theorem \ref{dedege} }

\begin{proof}
We look to the magnetic problem given by  the
2-form $w_{\epsilon}= \epsilon d\lambda + B \Omega$ for $H=\frac{1}{2}$ where
$\epsilon = \frac{1}{e}$. According to
table 1.1 the Hamiltonian vector field for $w_{\epsilon}$ is given
by a 
scaling of the Hamiltonian vector field for $w$. Thus their
characteristics
differ only by a time reparametrization. By theorem
\ref{pronn} there is a
neighborhood $N$ of $L_c$ and a diffeomorphism $\Xi_{\epsilon} : N
\rightarrow N$
such that
$$ \Xi_{\epsilon}^* {w_{\epsilon}} =
\gamma - (\epsilon^2 dB^{-1} + \epsilon^3 dF) \wedge d\theta
.$$
Thus the characteristic line bundle of $ \Xi_{\epsilon}^* {w_{\epsilon}}$
is spanned
by the Hamiltonian vector field given by the Hamiltonian system with
$$H_{\epsilon}=\epsilon^2 B^{-1} + \epsilon^3 F $$and
symplectic 2-form
$\Omega$ (where by abuse of notation we denote $\Xi_{\epsilon}^* \Omega $
by $\Omega$).
In section \ref{actanle} we introduced  action-angle
coordinates on the 
neighborhood of $\Xi^* N$ and
 reduced the dynamics of the Hamiltonian system
$(H_{\epsilon},\Omega)$
 on this neighborhood to the dynamics of a twist map. The
nondegeneracy
condition of Moser's twist theorem is calculated to be equation
\ref{dedege} according to (\ref{imag}) and lemma \ref{primeira}.
So Moser's twist
 theorem applies and invariant circles exist. By
dimensionality, they trap
the charge for all times $t > 0 $\end{proof}
\begin{proof} {\it (of Corollary \ref{nod1})} Observe that
each term in (\ref{deg}) depends on diferent powers of $|\nabla B|$. A 
careful analysis of this fact (done in section \ref{append}) implies that
(\ref{deg}) can not be satisfied near a critical point of $B$. So for a 
sufficiently large
 $e$ we can find a small neighborhood of $p$ where condition (\ref{dege})
is satisfied everywhere.\end{proof}  
 
\begin{proof} {\it (of Corollary \ref{nod2})} 
The proof is by contradiction. Suppose that given a neighborhood $N$
of $L_c$ we have that all the level sets contained
on $N$ are degenerate. This implies that exists a constant
$K$ such that 
$$ \frac{\partial^2 B^{-1}}{\partial^2 I}=K,$$
on $N$. This is equivalent to
$$-\frac{1}{B^2} \frac{\partial B}{\partial I}=\frac{-1}{B^2}
\frac{\partial}{\partial B}  =K.$$ What implies that
$$\frac{-1}{KB^2} = \frac{\partial I}{\partial B} = \int_{L_c} \frac
{B}{| \nabla B |} ds.$$
Since the right hand side is positive, the constant $K$ must be negative
and writing $ |K| = -K$ it follows that
$$ \frac{ |K|}{B^3} =\int_{L_c} \frac{1}{| \nabla B |} ds \ge \frac{l(c)}{
max\left(| \nabla B |\right)}.$$
where $l(c)$ denotes the length of the level $L_c$.
Thus we have that
\begin{equation}
 l(c) \le \frac{|K| max\left(| \nabla B |\right)}{m^3} .
\label{domina}
\end{equation}
where $m$ is the minimum of $B$ on $\overline N$, the closure of $N$.
 (\ref{domina}) implies that $l(c)$
is dominated by $max\left(|\nabla B(c)|\right)$.
Since $N$ was arbitrary this implies that $l(c)$ the length of the
critical level set must be zero, but this contradicts the hyphothesis. 
\end{proof}
\section{The Symmetric Case}
  In this section we deal with symmetric magnetic fields on a three
dimensional manifold, namely  ${\mathbb R}^3$ with the z-axis removed. Using
symplectic reduction
 the dimension of the system will be reduced allowing us to apply the main
 theorem of this work. The  non-degeneracy condition of one of the components
 of the magnetic field will be sufficient to apply Moser's Twist theorem. \par
  Let $M$ be the standard euclidean three dimensional space with the z-axis
 removed. Let $S^1$ act on ${\mathbb R}^3$ by rotation around the z-axis.
Let an
$S^1$ invariant 1-form $A$ on $M$ be given. Using cylindrical coordinates
 we write 
$$ A = A_r dr + A_{\theta} d\theta + A_z dz.$$
 The magnetic system given by
$ \beta = dA $ and the purely metric hamiltonian is equivalent to the
Hamiltonian system given by
$$ H = \frac{(p_{\theta} - e A_{\theta})^2 }{ 2 r^2} + \frac{(p_r - e
A_r)^2}{ 2} + \frac{(p_z - e A_z)^2 }{ 2}$$
and symplectic 2-form
$$ w = dp_{\theta} \wedge d\theta + dp_r \wedge dr + dp_z \wedge dz.$$
The momentum map for rotation around the z-axis is $p_{\theta}$. Fixing
$p_{\theta}=M$
it follows by a theorem of M. Kummer \cite{Kum} that the reduced
Hamiltonian is
$$ \hat H = \frac{p_r^2 }{ 2} + \frac{p_z^2 }{ 2} + V_{eff} $$  
where
$$ V_{eff}= \frac{(M - e A_{\theta})^2 }{ 2 r^2 },$$
the reduced symplectic 2-form  is given by
$$ \Omega = dp_r \wedge dr + dp_z \wedge dz + e \left(\frac{\partial A_z }{
\partial r} - \frac{ \partial A_r}{ \partial z} \right) dr \wedge dz$$
and the reduced manifold $N$  is given by  $N={\mathbb R} \times
{\mathbb R}^+$ with
coordinates $(z,r)$ which can be visualized as a half-plane with boundary
the z-axis. Writing $\beta$ as
$$ \beta = B_{\theta} dr \wedge dz + B_r d\theta \wedge dz + B_z dr \wedge
d\theta $$
we see that
$$ \Omega = dp_r \wedge dr + dp_z \wedge dz + eB_{\theta} dr \wedge dz=
d\lambda +eB_{\theta} dr \wedge dz $$
here $d\lambda$ is the canonical 1-form on $T^*N$. Assuming that
$$ E > V_{eff}  $$
we have that the  hypersurfaces
$$M_E= \left\{\hat H = E \right\}$$
and $$M_1= \left\{ \left(E - V_{eff}\right)^{-1} \left( \frac{p_r^2 }{ 2} +
\frac{ p_z^2 }{2} \right) = 1 \right\}$$
are equal. It follows that the closed characteristics of $\left.
 \Omega\right|_{M_E}$ are
equivalent to the closed characteristics of $\Omega |_{M_1}$. Thus we have
that
the reduced dynamics is (up to reparametrization) given by  the purely
 metric Hamiltonian
$$ {\bar H} =\frac{\left\| p \right\|_{g_E}^2 }{ 2}$$
(where $ p=(p_z,p_r)$ and $g_E$ is the metric
$  g_E = (E - V_{eff})^{-1} g$
where $g$ denotes the Euclidean metric on $T^*N$) and symplectic
2-form $\Omega$.
Assume that a level set of $B=(B_{\theta}, B_z,B_r)$ restricted to $N$
is  a simple closed curve, call it $L_c$,  and 
that for a neighborhood of $L_c$, the  non-degeneracy condition holds,
that is to say that $B_{\theta}$ satisfies (\ref{deg}) on a
neighborhood of $L_c$. Then we can apply Moser's twist theorem as in the 
proof of our main theorem.\par
 This result should be compared with the   
works of F. Truc \cite{Truc} and M. Braun \cite{Braun}. They considered
symmetric magnetic fields on
${\mathbb R}^3$ given by vector potentials of the type $ A_{\theta} dr \wedge
dz$, which is to say $B_{\theta}=0$. Those are extremely degenerate
magnetic fields in the sense of (\ref{deg}) and so our work does not
apply.
Also in their case, the absence
ot the $B_{\theta}$ component of their magnetic fields simplified the problem,
after the reduction of the symmetry, to a problem of the type  kinetic
 plus potential. Braun's work is, to the author's knowledge,
the first
 to apply Moser's twist theorem to a magnetic problem. Truc's work is
remarkable since she was able to prove the trapping of the particle
in the case where the magnetic moment was not convex. To our knowledge
we are the first to generalize the problem to a Riemmanian surface and to
deal with problems where the magnetic field could not be eliminated by
the reduction of the symmetry. It is in this sense that our work complement
theirs since we dealt with disjoint classes of magnetic field.

\section{Technical Proofs}
\label{append}
\begin{proof} {\it (of Corollary \ref{lema1})}  We follow McDuff
\cite{Mcduff}. The first part of
\ref{lema1} is an imediate consequence of Moser's homotopy argument.
The diffeomorphism  $\Phi_{\epsilon , \theta }$ is built by realizing it
as the flow of a vector field $X$ on $M$. First we consider the family of
2-forms
$w_t=(1-t)B\Omega + t \bar{w_{\epsilon}}$ for $t \in [0,1]$. We want a vector
 field $X$ such
that its flow  $\Phi_{\epsilon , \theta }$ is such that
 $$\Phi_{\epsilon , \theta,t }^* w_t = B \Omega.$$
 (in what follows we will omit the $\theta$ and $\epsilon$ dependency
 from the notation). Differentiating this relation we obtain that
$$ \frac{d}{ dt}\Phi_t^* w_t + \Phi_t^* \frac{ d w_t}{ dt}
 = 0, $$
i.e. that,
$$ \Phi^* \left(L_X w_t + \left( \bar{w_{\epsilon}} - B \Omega \right)
 \right)= 0. $$
Since $\Phi$ is a diffeomorphism and $w_t$ is closed it follows that
this can be satisfied if
\begin{equation}
 d(i_X w_t) + ( \bar{w_{\epsilon}} - B \Omega ) = 0.
\label{xx}  
\end{equation}
 We want to solve equation (\ref{xx}) for $X$. Since $w_t$ is
non-degenerate,
 it suffices to find a family of 1-forms $\sigma_t$ such that
\begin{equation}
 { \sigma_{t}}|_{TL_c} = 0, \hskip 0.3cm   d\sigma_1 = w_{\epsilon} -
 B \Omega .
\label{propp}
\end{equation}
In fact, conditions (\ref{propp})  imply 
\begin{equation}
 \sigma_{t} + i_{X_t} w_t =0 .
 \label{equaa} 
\end{equation}
 To construct  
$\sigma_t$ we consider the restriction of the exponential map to the
normal bundle $TL_c^{\perp}$ of the submanifold $L_c$ with respect to the
Riemannian metric on $M$. We denote this restriction by
\begin{equation}
 exp:TL_c^{\perp} \rightarrow M.
\label{nnno}  
\end{equation}
Consider the neighborhood of the zero section
$$U_{\epsilon} =\left\{ (m,v) \in TM \; | \;  m \in M ,\, v \in
TL_c^{\perp} ,
\, |v|
< \epsilon \right\}.$$
Then the restriction of the exponential map to $U_{\epsilon}$ is a
diffeomorphism onto $N_0=exp(U_{\epsilon})$ for $\epsilon > 0$
sufficiently small. Let $(p,v) \in TM$. Define $\pi_2$ as
$$\pi_2(p,v)= v.$$
Now define $\psi_t : N_0 \rightarrow N_0 $ for
$ 0 \le t \le 1$ by
$$\psi_t(q) = exp(q,tv), \quad\hbox{where} \quad v=\pi_2\left(exp^{-1}\left(q
\right) \right).$$
Then $\psi_t$ is a diffeomorphism for $t > 0$ and we have
$\psi_0(N_0) \subset L_c$, $\psi_1 = id$, and $\psi_t |_{L_c} = id.$
Calling $\tau = w_\epsilon - B \Omega$ we define
\begin{equation}
 \sigma_t \equiv \int_0^t \frac{d }{ dt} \psi^*_t \tau .
\label{defe}
\end{equation}
It's a direct computation to check that the family of 1-forms $\sigma_t$ has
the desired properties. Thus the family ($\sigma_t$) satisfy conditions
\ref{propp} and it follows
that $\Phi_{\epsilon , \theta}$ the flow of $X$, the vector field that
satisfies (\ref{equaa}) is such that
$$\Phi_{\epsilon , \theta}^* \bar{w_{\epsilon}} = B\Omega . $$
It follows that
$$\sigma_t=\int_0^t \frac{d }{dt} \psi_t^* (w_{\epsilon} - B\Omega)=
\int_0^t \frac{d}{dt}\psi_t^* \left\{ \epsilon^2( \frac{K}{2B} \Omega - \frac{
dB^{-1}}{ 2}) + \epsilon^3 R \right\} $$
where $R$ stands for the tail of the expansion of $\tau$. Denoting $ F = (
\frac{K }{ 2B} \Omega - \frac{ dB^{-1} }{ 2}) $ we have
$$\sigma_t=\epsilon^2 \int_0^t \frac{d }{dt} \psi^* F + \epsilon^3 \int_0^t
\frac{d }{dt} \psi^* R , $$
that we write as,
$$ \sigma_t = \epsilon^2 \sigma_{0,t} + \epsilon^3 \sigma_{1,t}  $$
where $  \sigma_{0,t} =\int_0^t \frac{d }{ dt} \psi^* F $ and $ \sigma_{1,t}
 =\int_0^t \frac{d }{dt} \psi^* R $. To solve equation (\ref{xx}) we
write
$$ X = \epsilon^2 X_{0,t} + \epsilon^3 X_{1,t},$$
where $X_{i,t}$ is the solution of
$$ \sigma_{i,t} + i_{X_{i,t}} w =0 . $$
for $i=0,1.$   We write $X = \epsilon^2 Y$ where $Y= X_{0,t} + \epsilon
X_{1,t}$. Since $X$ is a continuous vector field it follows
that $|Y|$ is bounded on a neighborhood of $N_0$ by a constant K.
The equation for the flow $\Phi$ is  
$$ \frac{ d \Phi }{ dt} = X(\Phi(t)).$$
Chosing $\Phi(0)=p_0 \in N_0$ and integrating it follows 
$$ \Phi(p_0,t) - \Phi(p_0,0) = \int_0^t X(\Phi(t)) dt .$$
Writing $\Phi$ the flow of $X$ as
$ \Phi= Id + \epsilon^2 C(\epsilon)$
we obtain 
$$ p_0 + \epsilon^2 C \left( p_0,\epsilon,t \right) - p_0 = \epsilon^2
\int_0^t Y(\Phi(t)) dt , $$
giving 
$$ |\epsilon^2 C(\epsilon)|
\le \epsilon^2 \int_0^1 |Y| dt \le \epsilon^2 K $$
and the result follows. \end{proof}
\begin{Remark} An important consequence of this lemma is that
$X_{0,t}$ is fiber independent. This follows directly from 
the definition. \end{Remark}

\begin{proof} {\it (of Corollary \ref{scct})} Using the Fermi coordinates
let $p=(x,y,\theta)$ be a point of $S^1M$.
 Omiting the $\epsilon$ dependence  of $\Xi$ for notational convenience we can
 write that
$$ \Xi (p) = (\Xi_x(p),\Xi_y(p),{\theta}).$$
The derivative of $\Xi$ can be calculated as
$$ \Xi_* = \left( \begin{array}{ccc}  \frac{\partial \Xi_x }{\partial x} &
\frac{\partial \Xi_x }{  \partial y}
 & \frac{\partial \Xi_x }{ \partial \theta}   \\
 \frac{\partial \Xi_y }{  \partial x}   &  \frac{\partial \Xi_y }{ \partial
y} & \frac{\partial \Xi_y }{ \partial \theta} \\  0     & 0  & 1
\end{array} \right)$$  which give us that
\begin{equation} \Xi_*(p) \frac{\partial }{ \partial \theta} = \Xi_*(p) \left(
\begin{array}{c}
 0 \\ 0 \\ 1 \end{array} \right) 
= \left( \begin{array}{c} \frac{\partial \Xi_x }{  \partial
\theta}(p\prime)
 \\ \frac{\partial \Xi_y }{\partial \theta}(p\prime) \\ 1 \end{array} \right),
\label{ster}
\end{equation}
where $p\prime= \Xi(p)$.
Recalling the definitions and lemma \ref{lema1} we can write that
$$ \left\{  \begin{array}{c}  \frac{\partial \Xi_x}{  \partial \theta}
=\epsilon^2
 \frac{\partial C_x }{\partial \theta} , \\ 
 \frac{\partial \Xi_y}{  \partial \theta} =\epsilon^2 \frac{\partial
C_y}{\partial \theta}.  \end{array} \right.$$

Since $\Phi_{\epsilon , \theta}= 1d +\epsilon^2 C(\epsilon,\theta)$ is the
flow of $X = \epsilon^2 X_{0,t} + \epsilon^3 X_{1,t} $ we have 
$$ \frac{d \Phi_{\epsilon , \theta} }{dt } = X(\Phi_{\epsilon , \theta}). $$
Integrating this equation we have that
$$ \epsilon^2 C(\epsilon, \theta) = \int_0^t \left(\epsilon^2 X_{0,1} +
 \epsilon^3 X_{1,t} \right)\circ \left(1_d +\epsilon^2 C(\epsilon,\theta)
\right)dt
,$$
that simplifies to,
\begin{equation}  
C(\epsilon, \theta) = \int_0^t \left\{ X_{0,t}(1_d + \epsilon^2 C(\epsilon
,\theta))+\epsilon X_{1,t} + \epsilon^3 X_{1,t}(C(\epsilon,\theta)) \right\}
dt.
\label{zde}
\end{equation}
Expanding $  X_{0,t}(1_d + \epsilon^2 C(\epsilon,\theta))$ in $\epsilon$,
inserting the resulting expression in (\ref{zde}) and collecting terms we
have 
$$ C(\epsilon, \theta) = \int_0^t  X_{0,t} dt + \epsilon \int_0^t X_R
dt,$$ 
for some $X_R$. Since $\, X_{0,t}$ is $\theta$ independent, we obtain 
$$ \frac{ \partial  C(\epsilon, \theta) }{ \partial \theta} =\epsilon 
\frac{\partial }{ \partial \theta} \int_0^t X_R dt.$$
 This is to say that
$ \frac{ \partial  C(\epsilon, \theta) }{ \partial \theta} $ is of order   
at least one in $\epsilon$ and we write that
$$  \frac{ \partial  C(\epsilon, \theta) }{ \partial \theta} = \epsilon
\bar C(\epsilon,\theta),$$
where
$$ \bar C(\epsilon,\theta)=\frac{\partial }{\partial \theta} \int_0^t X_R
dt.$$ Equation (\ref{ster}) gives 
$$ \Xi_*(p) \frac{\partial }{\partial \theta} = \Xi_*(p)
\left(\begin{array}{c} 0 \\ 0 \\ 1 \end{array} \right)=\left( \begin{array}{c}
\epsilon^3 \bar C_x(p\prime)
 \\ \epsilon^3 \bar C_y(p\prime) \\ 1 \end{array} \right) ,  $$
and we write 
$$  \Xi_*(p) \frac{\partial }{ \partial \theta} = \frac{\partial}{ \partial
\theta} + \epsilon^3 z, $$
where
$$ z =\left( \begin{array}{c} \bar C_x \\ \bar C_y \\ 0 \end{array}
\right), $$ proving the lemma. \end{proof}

\begin{proof} {\it (of Corollary \ref{nod1})} We will study each term 
of (\ref{deg}) individually.
Let $B(p)=a$. Since $p$ is a nondegenerate minimum (or maximum)
 point we have by Morse's lemma that there is a neighborhood $V$ and a
 system of coordinates $(u,v)$ on $V$ such that $B(u,v) = a \pm
\frac{ 1}{ 2} \left( u^2 + v^2 \right) $ where the
plus sign is used if the point is a minimum and the minus sign is used if the
point is a maximum.
\begin{Lemma}
Let $r_e^2=u^2 + v^2$ and let $r_g^2=|(u,v)|^2$. Then we have that
\begin{equation}
 |\nabla B|^2 = \frac{trace[g_{ij}] r_e^2 - r_g^2 }{ g }.
\label{fre}
\end{equation}
\end{Lemma}
The proof of this lemma is straightforward. Note that (\ref{fre}) implies
that $ trace[g_{ij}]r_e^2 - r_g^2 > 0 $ on
$V-\left\{p \right\}$. For a level set $c$ not equal to $a$ and such that
$L_c \subset V-\left\{p \right\} $ we have that $r_e$ is constant (since
B is) and we can write that
$$ g | \nabla B |^2 \ge trace[g_{ij}] r_e^2 ,$$

i.e.,
$$| \nabla B | \ge \left( \frac{trace[g_{ij}] }{ g} \right)^{\frac{1}{2}}
r_e. $$
Now we can see that the first term of (\ref{deg}) is bounded, in fact we
have
that
$$  \int_{L_c}  \frac{3 }{ |\nabla B|} fdx < 3
\left( \frac{ g }{ trace[g_{ij}] }\right)^{\frac{1}{2}}  \int_{L_c} f \ dx,
$$   but on $V$ we have that  $ g $ and $f$ are bounded above and
 $trace[g_{ij}]$ is bounded below. So we can find constants
$M$ and $K$  such that
$$ M \ge \left(\frac{ g }{ trace[g_{ij}] }\right)^{\frac{1}{2}}, \qquad 
and
 \qquad K \ge f  $$
on $V$, allowing us to write that
\begin{equation}
  \int_{L_c}  \frac{3}{ |\nabla B|} fdx < \frac{M K}{r_e} \int_{L_c}  dx
 =\frac{M K}{r_e} 2 \pi r_e = 2\pi M K.
\label{firsterm}
\end{equation}
This give us that the first term of (\ref{deg}) is bounded for
any level set close enough to $p$ as claimed. Now we look to the second term
\begin{equation}
  \int_{L_c} \frac{ B }{|\nabla B|^4} < \nabla |\nabla B|, \nabla B > fdx.
\label{dese}
\end{equation}
For a constant metric one  computes that
$$ < \nabla |\nabla B|, \nabla B >=|\nabla B|. $$
 (\ref{dese}) becomes
$$  \int_{L_c} \frac{ B }{ |\nabla B|^3} fdx,$$
considering lemma (\ref{fre}) , and proceeding as before we have that
\begin{equation}
  \int_{L_c} \frac{ B }{ |\nabla B|^3} fdx \ge \frac{ c}{ K} \int_{L_c}
\frac{f }{ r_e^3} dx \ge  \frac{ 2 \pi c }{ M} \frac{K }{ r_e^2};
\label{secondterm}
\end{equation}
The case of a nonconstant
 metric can be reduced to the case of a constant metric by choosing normal
coordinates on a neighborhood of
$p$. \par
 The last term of (\ref{deg}) can be minorated by noticing that
$$ \left\vert \int_{L_c} \frac{B }{|\nabla B|^3} <\nabla f, \nabla B> dx
\right\vert \le  \int_{L_c} 
\frac{B }{ |\nabla B|^2} |\nabla f| \le G \int_{L_c}
 \frac{1 }{ |\nabla B|^2},  $$

for some constant $G$ (since $ |\nabla f|$ and $B$ are bounded on $V$).
Now observing that
$ |\nabla B |^2 = r^2_g$, 
(\ref{fre}) implies 
$$ |\nabla B |^2 = \left( \frac{trace[g_{ij}] }{g + 1} \right) r_e^2. $$
And we can write 
$$\frac{1 }{|\nabla B |^2} \le \frac{H }{ r_e^2}$$
for some constant $H$ such that
$$  \frac{g+1 }{trace[g_{ij}]}\le H$$
on $V$. It follows that
\begin{equation}
\int_{L_c} \frac{B }{|\nabla B|^3} <\nabla f, \nabla B> dx  \le 2 \pi G
\frac{ H }{r_e}.
\label{thirdter}
\end{equation}
Now considering (\ref{firsterm}), (\ref{secondterm}), (\ref{thirdter}) we
see that
 (\ref{deg}) implies 
$$  M K \ge \frac{ c }{ M} \frac{K }{ r_e^2} -  G \frac{ H }{r_e}, $$
which can not be satisfied if $r_e$ is small enough. Thus for a sufficiently
small neighborhood of $p$  all the level sets are non-degenerate.
\end{proof}


\begin{thebibliography} {99}
\bibitem[AM]{Foundations} Abraham, A. and Marsden, J. 1978 {\it
Foundations of
Mechanics} (New York : Addison-Wesley)
\bibitem[Ar]{Arn} Arnold V I 1996 {\it Proc. Steklov Inst. Math.} {\bf
216} 3
 
\bibitem[Ar1]{Ar1} Arnold V I (ed) 1988 {\it Encyclopaedia of
Mathematical  
Science (Dynamical Systems III)} Vol 3 (Berlin : Springer-Verlag)
 
\bibitem[Ar2]{Ar2} Arnold V I 1986 {\it Russian Math. Surveys}
{\bf41} (6) 1-21

\bibitem[Ar3]{Ar3} Arnold V I 1988 {\it Lectures Notes in Math} {\bf 1346}
 
\bibitem[Br]{Braun} Braun M 1970 {\it J. Diff. Equations} {\bf 8} 294 
-49

\bibitem[CZ]{CZ} Conley C and Zehnder E. 1983 {\it Invent. Math.} {\bf 73} 33
\bibitem[Gi1]{Gi1} Ginzburg V:  Contact and Symplectic Geometry
 1996  edited by C. B. Thomas ( Cambridge University Press : New York)

\bibitem[Gi2]{Gi2}
Ginzburg, V. L., 
\emph{Funct. Anal. Appl.}, {\bf 21} (2) (1987), 100--106.


\bibitem[Gi3]{Gi3}
Ginzburg, V. L.,  {\em Math. Z.},{\bf 223} (1996), 397--409.

\bibitem[KN]{KN} Kobayashy S and Nomizu K 1963 {\it Foundations of
Differential Geometry} (New York : Interscience Publishers)

\bibitem[Ku]{Kum} Kummer M 1981 {\it Indiana Univ. Math. J.} {\bf30} 28

\bibitem[Li]{Little} Littlejohn R G 1979 {\it J. Math. Phys.} {\bf 20}
2445   

\bibitem[McD]{Mcduff} McDuff, D. and Salamon, D 1995 {\it Introduction to
Symplectic
Topology}  (Oxford University Press Inc : New York)

\bibitem[Mo]{Mont} Montgomery R 1995 {\it Comm. Math. Phys.} {\bf 168} 651
 
\bibitem[Mos]{Mos} Moser J 1962 {\it Nach. Akad. Wiss.
G\"ottingen: Math. Phys.}{\bf 1} 1


\bibitem[Tr]{Truc} Truc F 1996 {\it Ann. Inst. Henri Poincar\'e (Physique
 Theorique)} {\bf 64} 127


\end{thebibliography}
 \end{document}